\documentclass[12pt,a4paper]{article}
\usepackage{graphicx}
\usepackage{times}
\title{\bf Modelling the synchrotron emission from O-star colliding wind binaries}
\author{Delia Volpi$^1$ \\
\vspace{1cm}\\
\normalsize $^1$ Royal Observatory of Belgium, Ringlaan 3, 1180 Brussels, Belgium}
\date{\mbox{}}
\begin{document}
\maketitle
\pagestyle{empty}
%
%
\def\bull{\vrule height .9ex width .8ex depth -.1ex}
\makeatletter
\def\ps@plain{\let\@mkboth\gobbletwo
\def\@oddhead{}\def\@oddfoot{\hfil\tiny\bull\quad
``The multi-wavelength view of hot, massive stars''; 39$^{\rm th}$ Li\`ege Int.\ Astroph.\ Coll., 12-16 July 2010 \quad\bull}%
\def\@evenhead{}\let\@evenfoot\@oddfoot}
\makeatother
%
%
\def\beginrefer{\section*{References}%
\begin{quotation}\mbox{}\par}
\def\refer#1\par{{\setlength{\parindent}{-\leftmargin}\indent#1\par}}
\def\endrefer{\end{quotation}}
%
%
{\noindent\small{\bf Abstract:} 
Many early-type stars are in binary systems. A number of them shows radio emissivity with periodic variability. This variability is associated with non-thermal synchrotron radiation emitted by relativistic electrons. The strong shocks necessary to accelerate the electrons up to high energies are produced by the collision of the radiatively-driven stellar winds. A study of the non-thermal emission is necessary in order to investigate O-star colliding wind binaries. Here preliminary results of our modeling of the colliding winds in Cyg~OB2 No.~9 are presented.
}
%
%
\section{Introduction}
Many OB stars show non-thermal radio emission together with the thermal free-free radiation in the same band. The non-thermal emissivity is thought to be due to synchrotron radiation emitted by relativistic electrons spiraling around the dipole magnetic field lines of the star. The electrons are accelerated up to high energies by strong shocks.

In recent years a number of early-type stars radiating non-thermal radio emission have been discovered to be binary systems. This discovery supports the hypothesis of colliding winds from the primary and the secondary as the origin of the strong shocks (see Eichler \& Usov 1993, Dougherty et al. 2003, Pittard et al. 2006). The observed synchrotron emissivity is characterized by a periodic variability directly connected with the orbital phase of the binary system.

\begin{figure}[th]
\centering
\includegraphics[width=8cm]{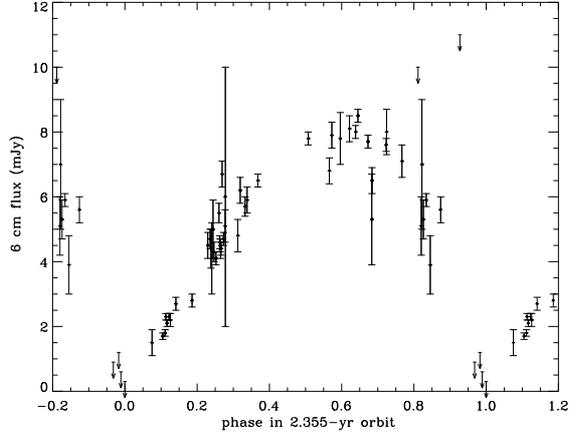}
\caption{Cyg~OB2 No.~9. Observed VLA $6$~cm radio emission (from Van Loo et al.\ 2008).\label{fig_1}}
\end{figure}

One of these binary systems is Cyg~OB2 No.~9, which was detected as a non-thermal emitter in 1984 (see Abbott et al.\ 1984). By studying the observed VLA (Very Large Array) radio fluxes at $3.6$, $6$ and $20$~cm (Van Loo et al.\ 2008) from Cyg~OB2 No.~9, Van Loo and collaborators found a variability with a period of $2.35$~yr. This periodicity confirmed the presence of radio non-thermal emission suggesting also a binarity of this system. At the same time Naz\'e et al.\ (2008) discovered its binarity after a long-term spectroscopic monitoring. The observed $6$~cm radio flux is shown in Fig.\,\ref{fig_1}. The orbital parameters of Cyg~OB2 No.~9 were derived for the first time by Naz\'e et al.\ (2010).

 The study of the synchrotron emission at different orbital positions is relevant to constrain stellar, orbital and wind parameters, such as the mass loss rates from the stars and the clumping/porosity, and in general to investigate the nature of the stellar winds. As a matter of fact the mass loss rate determination in massive stars is presently one of the main problems. This problem has become relevant due to the clumping/porosity of the stellar winds which affects the estimation of the mass loss rate. To determine the emitted synchrotron radiation and its absorption by the stellar wind plasma permits to estimate the amount of porosity in the wind.

In order to better understand the physics of the non-thermal emission and to constrain the model parameters we developed a numerical code. Here the simulated model for Cyg~OB2 No.~9 is presented. The theoretical results are compared with the corresponding observations.



\section{Modeling}
In this section the theoretical model used to obtain the simulated results is described. A first version of this model was developed by R.Blomme. The details of the model can be found in Blomme et al.\ (2010).

The two winds are considered to collide at the contact discontinuity. Its position is assumed to be coincident with the two shocks and it is calculated using the Anthokin et al.\ (2004) equations. The electrons accelerated at the shock are followed as they advect away and cool down along the post-shock streamlines. We stop following the electrons when their momentum falls below a minimum value (see Van Loo et al.\ 2004), or when they leave the simulation volume. Adiabatic and inverse Compton losses are both taken in account. The momenta follow a modified power-law distribution. 

The synchrotron emissivity from the relativistic electrons is calculated along the post-shock streamlines in the orbital plane. The Razin effect is also included. The formula of the emissivity $j_{\nu}(r)$ is reported below as in Van Loo et al.\ (2005):

\begin{equation}
j_{\nu}(r) = \frac{1}{4 \, \pi} \int_{p_{0}}^{p_{\mathrm{c}}} dp\,N(p,r) \int_{0}^{\pi}\frac{d \theta}{4 \pi} 2 \pi \sin{\theta} \frac{\sqrt{3}\,\mathrm{e^{3}}}{\mathrm{m_{e}\,c^{2}}}B\,f(\nu,p)\sin{\theta}\,F \left(\frac{\nu}{f^{3}(\nu,p)\,\nu_{\mathrm{s}}(p,r)\sin{\theta}}\right)
\end{equation}
where $\nu$ is the frequency, $\mathrm{r}$ is the radial position, $p_{0}$ is the minimum momentum, $p_{\mathrm{c}}$ is the maximum one, $p$ is the particle momentum, $N(p,r)$ is the modified power-law distribution of the electrons, $\theta$ is the pitch angle, $\mathrm{e}$ and $\mathrm{m_{e}}$ are the electron charge and mass, $c$ is the velocity of light, $B$ is the magnetic field at a distance $r$, $\nu_{\mathrm{s}}$ is the critical frequency, and $f(\nu,p)$ takes into account the Razin effect.

The synchrotron emission is calculated in the orbital plane. The third dimension is recovered by rotating the orbital plane along the line which connects the two stars. The free-free opacity and emission due to the ionized wind material are also included using the Wright \& Barlow (1975) equations. The theoretical fluxes are obtained at different orbital phases using Adam's method (Adam 1990) that solves the radiative transfer equation in a very simple way.

The orbital parameters are provided by Naz\'e et al.\ (2010) observations, the stellar parameters by Martins et al.\ (2005) theoretical models and the wind parameters by Vink et al.\ (2001) theoretical models. We choose spectral type O5I+O6I: the Martins' stellar parameters for this spectral type are consistent with the spectroscopic observations and the chosen spectral type is in agreement with the O5+O6-7 spectral types suggested by Naz\'e et al.\ (2008). Current spectroscopic information does not allow a reliable determination of the luminosity class.

The knowledge of the spectral type is necessary to select the values of parameters that cannot be obtained by the observations. These parameters are used as input to our model. The chosen values of the parameters are reported in Table\,\ref{parameter}.

\begin{table}
\caption{Stellar, orbital and wind parameters for Cyg~OB2 No.~9.}
\label{parameter} 
\small
\begin{center} 
\begin{tabular}{ |l |cc|}
\hline 
& primary star & secondary star \\
\hline
P~(days) & \multicolumn{2}{|c|}{860.2 $\pm$ 5.5}\\
e &  \multicolumn{2}{|c|}{0.744 $\pm$ 0.030}\\
i~(deg) & \multicolumn{2}{|c|}{48.59 $\pm$ 7.45}\\
$\omega$~(deg) & \multicolumn{2}{|c|}{-164.4 $\pm$ 4.1}\\
d~(pc) & \multicolumn{2}{|c|}{1820 $\pm$ 200}\\
$\nu_{\mathrm{0}}$~(km/s) & \multicolumn{2}{|c|}{0.1}\\
$M_{\mathrm{\star}}$~$(M_{\mathrm{sun}})$  &  50.72   &  44.10\\
$a~(R_{\mathrm{sun}})$        &  797.34 $\pm$ 203.95 &  932.40 $\pm$  237.17\\
$R~(R_{\mathrm{sun}})$                 &  19.45           &  19.95\\
$L$~$(\mathrm{erg}~\mathrm{s}^{-1})$  & 2.85$\cdot10^{39}$ &  2.48$\cdot10^{39}$\\
 $T_{\mathrm{eff}}$~(K)                &  38612              &  36801\\
 $v_{\infty}$~(km/s)                  &  2050              &  1850\\
 $\dot{M}$~$(M_{\mathrm{sun}}/yr)$    &  5.79d-06      &  4.86d-06\\
\hline 
\end{tabular}\\ 
\vspace{0.2cm}
 \parbox{15cm}{\footnotesize{\bf Note.} $P$ and $d$ are the orbital period and the distance from the Earth (Van Loo et al.\ 2008), $e$ and $\omega$ are the eccentricity and the angle of the periastron (Naz\'e et a.\ 2010), $M_{\mathrm{\star}}$, $R$, $L$, $T_{\mathrm{eff}}$ are the mass, the radius, the luminosity, and effective temperature for the star (Martins et al.\ 2005), $i$ is the inclination angle (obtained from a comparison between the theoretical mass $M_{\mathrm{\star}}$ calculated in Martins et al.\ 2005 and the projected observed one $M_{\mathrm{\star}}\sin{i}^{3}$ from Naz\'e 2010), $a$ is the semi-major axis of the orbit, $\nu_{0}$ is the velocity of the wind at the surface of the star (Blomme et al.\ 2010), $v_{\infty}$ and $\dot{M}$ are the velocity of the wind at large distance from the star and the mass loss rate (Vink et al.\ 2001).}
\end{center}
\end{table} 


\begin{figure}[t]
\centering
\includegraphics[width=18cm]{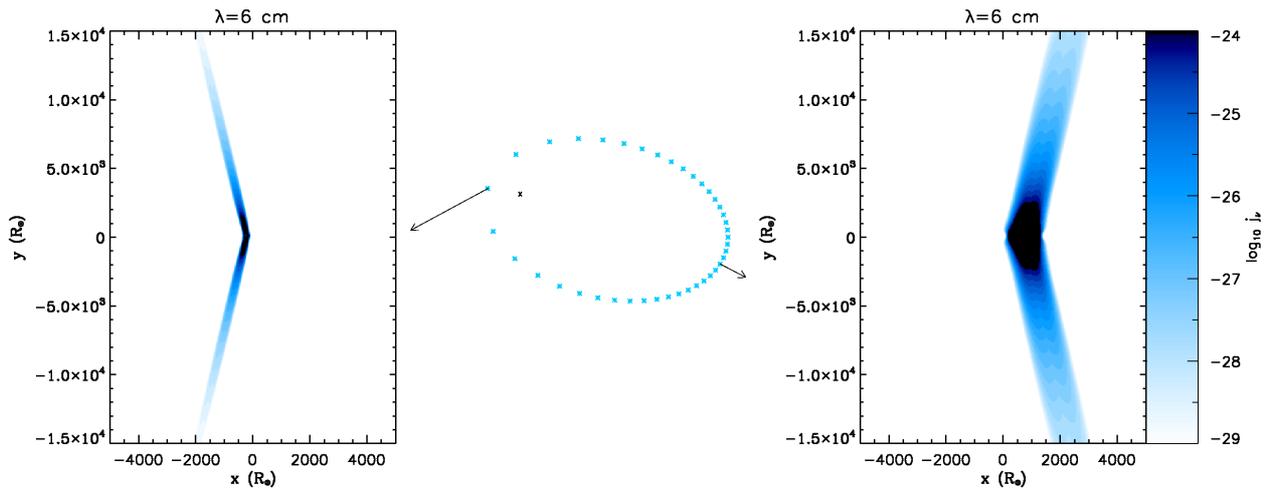}
\caption{Cyg~OB2 No.~9 simulated results. Left panel: the $6$~cm emissivity at periastron. Middle panel: the eccentric orbit of the binary system. Right panel: the $6$~cm emissivity at apastron. Along the axis the distances from the apex are in unit of $\mathrm{R_{Sun}}$. The color bar represents the emissivity in $\mathrm{erg/cm^{3}~s~Hz}$ with a logarithmic scale. The arrows link respectively the periastron and apastron positions with the corresponding emissivity pictures.\label{fig_2}}
\end{figure}

\begin{figure}[bh]
\centering
\includegraphics[width=8cm]{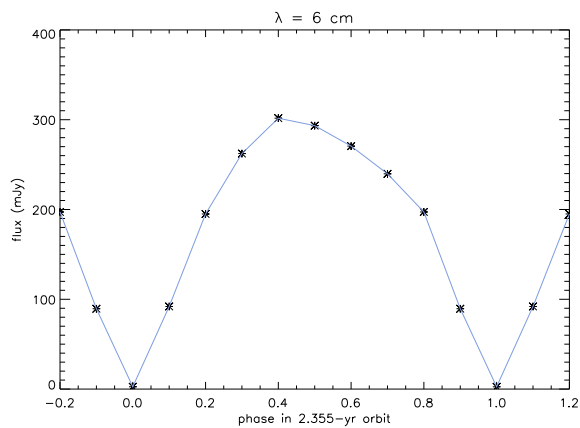}
\caption{Cyg~OB2 No.~9. Simulated $6$~cm light curve.\label{fig_3}}
\end{figure}

\section{Theoretical results: the emissivity and model light curve at $6$~cm}

The $6$~cm emissivity images at periastron and apastron are shown in Fig.\,\ref{fig_2}. The $6$~cm emissivity at periastron is much more concentrated around the contact discontinuity than at apastron mainly because the Razin effect is more important in the periastron high-density region. The synchrotron radiation is however not negligible in both cases due to the extension of the emitted area. This extended region explains the synchrotron emission that was observed in OB stars systems during the last years. 


In Fig.\,\ref{fig_3} the preliminary $6$~cm light curve from our modeling is presented. The variability in the radio flux is clearly visible as observed with VLA. This variability linked to the orbital period is the fingerprint of the non-thermal radiation. Compared to the observations (see Fig.\,\ref{fig_1}), the theoretical fluxes are much too high and the maximum occurs too early. This overestimate of the flux could be produced by too high a number of particles being taken into account. In order to reduce the theoretical fluxes we could decrease the fraction of energy transferred from the shock to the relativistic particles. In the present work the assumed value was the typical $0.05$ accepted for a strong shock following Eichler \& Usov (1993). Another option could be to increase the index $n$ of the $p^{-n}$ particle momentum distribution function and thus to decrease the shock strength, or even to diminish the star surface magnetic fields that we assumed to be $100$~Gauss. Before tuning these model parameters it is however more urgent to evolve directly the hydrodynamical equations in order to correctly position the shocks and obtain a realistic value for their strengths, and to take into account the instabilities and inhomogeneities that influence both the thermal and non-thermal emission. Then it would be relevant to include the orbital motion. 

Further investigation and changes in the model will be thus necessary to improve quantitatively the simulated results. 

\section{Future work}
Future work will consist first of all in solving directly the hydrodynamical equations and including the orbital motion.
Then  it will be necessary to explore the parameter space of our model to obtain a better agreement with the corresponding observations (included the stellar spectral types).
%
%
\section*{Acknowledgments}
This research is supported by contract Action 1 project MO/33/024 (Colliding winds in O-type binaries). 
%
%
\footnotesize
\beginrefer

\refer Abbott D.C., Bieging J.H., Churchwell E., 1984, ApJ, 280, 671

\refer Adam J., 1990, A \& A, 240, 541

\refer Anthokin I.I., Owocki S.P., Brown J.C., 2004, ApJ, 611, 434

\refer Blomme R., De Becker M., Volpi D., Rauw G., 2010, A \& A, 519A.111B

\refer Dougherty S.M., Pittard J.M., Kasian L., Coker R.F., Williams P.M., Lloyd H.M., 2003, A \& A, 409, 217 

\refer Eichler D., Usov V., 1993, ApJ, 402, 271

\refer Martins F., Schaerer D., Hillier D.J., 2005, A \& A, 436, 1049

\refer Naz\'e Y., De Becker M., Rauw G., Barbieri C., 2008, A \& A, 483, 543

\refer Naz\'e Y., Damerdji Y., Rauw G., Kiminki D.C., Mahy L., Kobulnicky H.A., Morel T., De Becker M., Eenens M., Barbieri P., 2010, ApJ, 719, 634

\refer Pittard J.M., Dougherty S.M., Coker R.F., O'Connor E., Bolingbroke N.J., 2006, A \& A, 446, 1001

\refer Van Loo S., Runacres M.C., Blomme R., 2004, A \& A, 418, 717

\refer Van Loo S., 2005, PhD Thesis, KULeuven,Leuven, Belgium, http://hdl.handle.net/1979/53(VL)

\refer Van Loo S., Blomme R., Dougherty S.M., Runacres M.C., 2008, A \& A, 483, 585

\refer Vink J.S., de Koter A., Lamers H.J.G.L.M., 2001, A \& A, 369, 574

\refer Wright A.E., Barlow M.J., 1975, MNRAS, 170, 41


\endrefer           
\end{document}